\documentclass[10pt,letterpaper]{article}
\usepackage[top=0.85in,left=2.75in,footskip=0.75in,marginparwidth=2in]{geometry}

\usepackage[utf8]{inputenc}

\usepackage{natbib}

\usepackage{nameref,hyperref}

\usepackage[right]{lineno}

\usepackage{amsmath}

\usepackage{comment}

\usepackage{microtype}
\DisableLigatures[f]{encoding = *, family = * }

\raggedright
\usepackage[colorinlistoftodos]{todonotes}

\usepackage{changepage}

\usepackage[aboveskip=1pt,labelfont=bf,labelsep=period,singlelinecheck=off]{caption}

\makeatletter
\renewcommand{\@biblabel}[1]{\quad#1.}
\makeatother

\usepackage{lastpage,fancyhdr,graphicx}
\usepackage{epstopdf}
\fancyheadoffset[L]{2.25in}
\fancyfootoffset[L]{2.25in}

\usepackage{color}

\definecolor{Gray}{gray}{.25}

\usepackage{graphicx}

\usepackage{sidecap}

\usepackage{wrapfig}
\usepackage[pscoord]{eso-pic}
\usepackage[fulladjust]{marginnote}
\reversemarginpar

\begin{document}
\vspace*{0.35in}


\begin{flushleft}
{\Large
\textbf\newline{$k$-mer-based approaches to bridging pangenomics and population genetics}
}
\newline
\\
Miles D. Roberts\textsuperscript{1$\dagger$*},
Olivia Davis\textsuperscript{2$\dagger$},
Emily B. Josephs\textsuperscript{4,5,6},
Robert J. Williamson\textsuperscript{2,3},
\\
\bigskip
\bf{1} Genetics and Genome Sciences Program, Michigan State University, East Lansing, MI 48824, USA
\\
\bf{2} Department of Computer Science and Software Engineering, Rose-Hulman Institute of Technology, Terre Haute, IN 47803, USA
\\
\bf{3} Department of Biology and Biomedical Engineering, Rose-Hulman Institute of Technology, Terre Haute, IN 47803, USA
\\
\bf{4} Department of Plant Biology, Michigan State University, East Lansing, MI 48824, USA
\\
\bf{5} Ecology, Evolution, and Behavior Program, Michigan State University, East Lansing, MI 48824, USA
\\
\bf{6}, Plant Resilience Institute, Michigan State University, East Lansing, MI 48824, USA
\\
\bigskip
$\dagger$ co-first authors

* corresponding author: milesdroberts@gmail.com

\end{flushleft}


\section*{Abstract}

Many commonly studied species now have more than one chromosome-scale genome assembly, revealing a large amount of genetic diversity previously missed by approaches that map short reads to a single reference. However, many species still lack multiple reference genomes and correctly aligning references to build pangenomes is challenging, limiting our ability to study this missing genomic variation in population genetics. Here, we argue that $k$-mers are a crucial stepping stone to bridging the reference-focused paradigms of population genetics with the reference-free paradigms of pangenomics. We review current literature on the uses of $k$-mers for performing three core components of most population genetics analyses: identifying, measuring, and explaining patterns of genetic variation. We also demonstrate how different $k$-mer-based measures of genetic variation behave in population genetic simulations according to the choice of $k$, depth of sequencing coverage, and degree of data compression. Overall, we find that $k$-mer-based measures of genetic diversity scale consistently with pairwise nucleotide diversity ($\pi$) up to values of about $\pi = 0.025$ ($R^2 = 0.97$) for neutrally evolving populations. For populations with even more variation, using shorter $k$-mers will maintain the scalability up to at least $\pi = 0.1$. Furthermore, in our simulated populations, $k$-mer dissimilarity values can be reliably approximated from counting bloom filters, highlighting a potential avenue to decreasing the memory burden of $k$-mer based genomic dissimilarity analyses. For future studies, there is a great opportunity to further develop methods to identifying selected loci using $k$-mers.

\section*{Introduction}

Two decades ago, assembling one reference genome for one eukaryotic species was an international, herculean effort \citep{venter_sequence_2001}. Now, individual laboratories are readily assembling pangenomes that include the alignment and comparison of multiple reference-quality genomes within the same species  \citep{golicz_pangenomics_2020}. These pangenomic analyses have uncovered vast quantities of genetic variation previously missed by the ubiquitous practice of aligning short reads to a single reference genome \citep{wong_towards_2020, ebler_pangenome-based_2022, zhou_graph_2022, rice_pangenome_2023, liao_draft_2023}. Pangenomes have also revised our understanding of previously cataloged variation, because single nucleotide polymorphisms once identified by mapping reads against a single reference, can sometimes represent structural variation \citep{jaegle_extensive_2023}. This better ability to capture variation, combined with the recent explosion of pangenomes for many non-model organisms \citep{lei_plant_2021}, means that pangenomic analysis will likely be a standard population genetics practice in the near future.

However, analyzing pangenomes is computationally intensive and remains an important open challenge in population genetics \citep{hahn_molecular_2018}. The key step in the construction of pangenomes is typically multiple sequence alignment (MSA) and performing MSA on many genomes is computationally intensive, especially if the genomes are large, repetitive, and complex \citep{zielezinski_alignment-free_2017, aylward_pankmer_2023, song_new_2024}.Even when given adequate computing resources, finding optimal MSAs is an NP-hard problem, meaning computing the optimal solution relatively quickly (i.e. in polynomial time) is almost certainly mathematically impossible \citep{WANG_complexity_1994, Just_computational_2001, Elias_settling_2006, Caucchiolo_hardness_2023}. 
MSA algorithms often rely on heuristics to derive approximate solutions and, even so, may not scale well enough to align many large, complex genomes at once \citep{song_new_2024}. Although modern MSA algorithms are impressive, a researcher's exact choice of alignment parameters and software can drastically affect downstream genotype calls \citep{Orawe_low_2013, Li_toward_2014, bush_genomic_2020, Betschart_comparison_2022, sopniewski_estimates_2024}. Aligning new sequences to pangenomes in order to call genotypes is also more intensive than common single-reference genotyping \citep{rice_pangenome_2023}, presumably because pangenomes contain alternative alleles and additional DNA not found in every reference, substantially increasing the number of possible alignments to sift though \citep{chin_multiscale_2023}. Finally, the difficulty of accurately aligning sequences increases with divergence \citep{bush_genomic_2020}. Thus, if the sequences one wants to compare through alignment are highly diverse or deeply diverged, there comes a point where producing an alignment that's better than random is difficult \citep{kille_multiple_2022}. Given these burdens of computing alignments and refining alignment parameters, approaches that skip MSA altogether could thus provide valuable complementary or alternative approaches to studies in population genetics. 

In this review, we argue that one concept that can facilitate alignment-free approaches, but needs more attention from population geneticists, is the $k$-mer. A $k$-mer is a sub-sequence of length $k$ within a larger sequence. The main benefit of $k$-mers is that they can be analyzed without alignment. Instead, one simply counts all of the $k$-mers present in a sample of reads, then uses the resulting count matrix for downstream analysis (see Figure \ref{fig:kvs}). Thus, $k$-mers derived from regions far diverged, absent, or otherwise unalignable in a given reference will not be automatically excluded from an analysis, allowing one to get a picture of pangenomic variation. 
$k$-mers have a long history of use in metagenomics \citep{mcclelland_selection_1985, rosen_metagenome_2008, dubinkina_assessment_2016}, phylogenetics \citep{kolekar_alignment-free_2012, Haubold_alignment-free_2014, zielezinski_alignment-free_2017, zielezinski_benchmarking_2019, bussi_large-scale_2021, beichman_evolution_2023, jenike_guide_2024}, computer science \citep{shannon_mathematical_1948}, and quantitative genetics \citep{voichek_identifying_2020, kim_dissecting_2020, Gupta_gwas_2021, Onetto_population_2022, lemane_kmdiff_2022}. However, while applications of $k$-mers receive much attention in other disciplines, population genetic investigations using $k$-mers remain limited. 

Our goal is to review $k$-mer-based approaches for identifying, measuring, and explaining patterns of genetic variation in populations. At the same time, we investigate the behavior of $k$-mer-based measures of variation to the choice of $k$, the depth of sequencing coverage, and the degree of data compression. We finally highlight some avenues to explore in $k$-mer-based (i.e. reference-free or alignment-free) population genetics. Overall, we advocate that $k$-mer-based approaches can be valuable complements to common reference-based or SNP-based population genetics methods.

\section*{Box 1: What is the ``best'' value for $k$?}

A common first question in $k$-mer based analyses is: What value(s) of $k$ should be analyzed?  The ``best'' $k$ for an analysis is ultimately determined by a trade-off between the length of $k$ and sequencing error: longer $k$-mers are more likely to represent unique genomic sequences, but are also more likely to contain a sequencing error \citep{rahman_association_2018}. In practice, many studies use a $k$ of around 20-40 bp \citep{ponsero_comparison_2023} because $k$-mers in this range can be reliably sequenced with short read data and often align uniquely to their source genome \citep{wu_effect_1991,becher_measuring_2022}. For example, $k = 32$ captures 85.7 \% of unique sequences in the human genome \citep{shajii_fast_2016}, while $k = 21$ distinguishes many eukaryote, bacteria, and archea species \citep{bussi_large-scale_2021}. However, there are many past studies that propose criteria for choosing specific values of $k$.

\subsection*{Choosing multiple values of $k$}

One ``brute force'' approach to test the sensitivity of results to the choice of $k$ is to simply repeat an analysis multiple times for different values of $k$. This approach is especially common among genome assembly algorithms \citep{chikhi_informed_2014, durai_informed_2016} but could be applied to almost any analysis in theory. The main drawback, however, is the high computational burden of performing the same analysis multiple times. It is also difficult to know without more information whether analyses performed for certain values of $k$ produce more accurate results than analyses for other values of $k$, motivating the need for $k$ selection criteria that can be either minimized or maximized.  

\subsection*{Choosing $k$ based on the number of unique non-erroneous $k$-mers}

Higher values of $k$ generally allow greater detection of unique sequences, but increase the probability of observing $k$-mers containing at least one sequencing error \citep{chikhi_informed_2014, rahman_association_2018}. Each sequencing error can result in up to $k$ erroneous $k$-mers, making it important to prevent errors from dominating one's analysis. Thus, choosing a $k$ that maximizes the number of unique non-erroneous $k$-mers in a dataset is generally considered optimal for tasks like genome assembly \citep{chikhi_informed_2014}. This approach generally involves measuring the copy number distribution of $k$-mers in a sample (i.e. the $k$-mer frequency spectrum or abundance histogram) and then fitting a model to the distribution to estimate which parts of the spectrum come from erroneous $k$-mers \citep{chikhi_informed_2014}. Usually,  the low-frequency end of the spectrum is dominated by erroneous $k$-mers because sequencing errors are unlikely to generate the same erroneous $k$-mers many times and usually convert real $k$-mers into $k$-mers not found in the source genome \citep{kelley_quake_2010}. Although this criterion for choosing $k$ could be applied to population genetic datasets, it is not if clear the resulting optimal $k$ would vary significantly between genomes within the same species. Presumably, if genomes within the same species have considerable variation in repetitive content \citep{haberer_european_2020}, size \citep{schmuths_genome_2004}, or ploidy (reviewed in \citet{kolar_mixed-ploidy_2017}) then the optimal $k$ could vary. Determining the value of $k$ that maximizes the number of unique non-erroneous $k$-mers across all individuals in a population may be of interest for future population genetics studies.

\subsection*{Choosing $k$ based on a desired false-positive rate}

The ability of $k$-mers in the range of short-read technology (20 - 40 bp) to differentiate genomes is also supported mathematically. Here we present equations similar to ones in \citet{ondov_mash_2016}, except we generalize to account for variation in GC content between sequences being compared. We start by defining the probability of a $k$-mer $K$ randomly sampled from a genome sequence $X_1$ being found in a genome sequence $X_2$ of size $L$ in base pairs as:

\begin{equation} \label{kfp_complex}
    P(K \in X_2) = 1 - \left(1 - \Sigma^k\right)^{(L + 1 - k)}
\end{equation}
where $\Sigma$ gives the probability of a chance match between $K$ and $X_2$ at any single position. If $X_1$ and $X_2$ contain equal proportions of all bases, then $\Sigma = \frac{1}{4}$ (the probability of drawing both an A, T, G, or C from $X_1$ and $X_2$). If we generalize to allow the proportion of some base in $X_1$ (it does not matter which) to be an arbitrary value $b_1$ and the proportion of the same base in $X_2$ to be an arbitrary value $b_2$, then $\Sigma = 4b_1b_2 - b_1 - b_2 + \frac{1}{2}$ where $b_1,b_2 \in [0, \frac{1}{2}]$ (obtained by summing up the probabilities of drawing both an A, T, G, or C from $X_1$ and $X_2$).  Assuming $k << L$, Equation \ref{kfp_complex} approximates to:

\begin{equation} \label{kfp_approx}
    P(K \in X) \approx 1 - \left(1 - \Sigma^k\right)^{L}
\end{equation}
As in \citet{fofanov_how_2004}, we can then solve Equation \ref{kfp_approx} to give the minimum $k$-mer length required to achieve a desired false positive rate of $q = P(K \in X)$:

\begin{equation} \label{eq:kfp_final}
    k = \left \lceil log_{\Sigma}\left(1 - (1 - q)^{1/L}\right) \right \rceil
\end{equation}

Equation \ref{eq:kfp_final} demonstrates that $k = 19$ reduces the probability of two 3 Gb genomes of random sequence sharing the same $k$-mer by chance to just 1 \% \citep{ondov_mash_2016}  (assuming all bases are in equal proportion in both genomes, $b_1=b_2=\frac{1}{4}$) while $k = 27$ gives $q = 1 \times 10^{-6}$ for a 10 Gb genome where both $X_1$ and $X_2$ have a GC conent of 42 \% (i.e. $b_1,b_2 = 0.21$). Slightly longer $k$-mers are required when the proportion of each base is not 25 \% in both genomes, reflecting the fact that spurious matches are more likely when the genomes are biased toward certain bases. For example, the required $k$-mer length increases to 34 bp if the GC content of both genomes is 21 \% ($b_1,b_2 = 0.105$), given a 10 Gb genome and desired false positive rate of $q = 1 \times 10^{-6}$.

In summary, if $k$ is high enough, two genomes in a population are highly unlikely to share $k$-mers just through the accumulation of random sequences alone, suggesting that shared $k$-mers usually reflect shared ancestry. Developing additional $k$-mer selection criteria that explicitly model the influence of shared ancestry on the probability of $k$-mer sharing could improve approaches to choosing $k$.

\subsection*{Choosing $k$ based the balance of shared vs differing $k$-mers}

A final class of approaches for choosing $k$ focuses on the following central idea \citep{zhang_viral_2017}: when $k$ is small, most $k$-mers are shared between most samples, but as $k$ increases more sample-specific $k$-mers occur until no more shared $k$-mers remain. Neither of these scenarios is usually desirable - the former makes all samples appear to be identical, while the later makes all samples appear to be completely different. Thus, choosing $k$ boils down to balancing the number of shared vs differing $k$-mers. Some statistics for informing this choice include cumulative relative entropy \citep{sims_alignment-free_2009}, relative sequence divergence \citep{sims_alignment-free_2009}, average number of common features \citep{zhang_viral_2017}, Shannon's diversity index \citep{zhang_viral_2017}, and $\chi^2$ tests \citep{bai_optimal_2017}. However, there are two main drawbacks of these statistics. First, they were mainly developed for phylogenetic studies focused on short amino acid $k$-mers, so their utility for population genetics (which would probably focus on longer nucleotide $k$-mers) is untested. Furthermore, different values of $k$ will appear optimal for genomes of different sizes \citep{zhang_viral_2017}, making it hard to imagine that one optimal $k$ exists for a population exhibiting substantial genome size variation. Extending these or similar approaches to optimize $k$ for populations with genome size variation will be very useful for future studies.

\section*{Box 2: What's the expected value of $k$-mer diversity in a neutrally evolving population?}

A key question of interest for population geneticists is the expected value of $k$-mer metrics for populations evolving under a given model. Here, we derive a simple bound on the expected number of $k$-mer differences between a pair of individuals in a neutrally evolving population. Our result is similar to \citep{shi_alignment-_2024}, except we generalize beyond haploid organisms.

We start by assuming that (1) $k$ is sufficiently long to capture all of the unique sequences in a genome, (2) each genome is sequenced at sufficient coverage to confidently identify all of the $k$-mers present, (3) $k$-mers are sequenced without error, and (4) only SNPs contribute to differences between genomes. Let $i$ and $j$ be individuals in a neutrally evolving population with ploidy level $x$ and let $y$ and $z$ be two haploid genomes in the pool of $i$ and $j$ that differ by origin. Next, let $K_i$ and $K_j$ represent the set of $k$-mers identified in individual $i$ and $j$'s genomes respectively and let $|K_i|$ denote the size of the set $K_i$. If all of the genomes within $i$ and $j$ are identical, then all $k$-mers are shared between them. If we start by assuming $i$ and $j$ are haploid, this means a maximum of $2k$ $k$-mers are not shared between $i$ and $j$ for every pairwise difference between $i$ and $j$ \citep{iqbal_novo_2012, younsi_using_2015}. This relationship can be written as:

\begin{equation} \label{eq:haploid_diff2k}
    |K_i \cup K_j| - |K_i \cap K_j| \leq 2kD_{yz}
\end{equation}

where $D_{yz}$ represents the number of pairwise differences between haplotypes $y$ and $z$ and the left side represents the number of $k$-mers exclusive to either $i$ or $j$. The reason for the using a ``$\leq$'' in Equation \ref{eq:haploid_diff2k} is to account for three possibilities: (1) if two SNPs are less than $k$ base pairs apart, there will be fewer than $4k$ $k$-mers not shared between $i$ and $j$, (2) it's possible for a SNP to turn one $k$-mer into a different $k$-mer that's already present elsewhere in a genome and (3) at higher ploidy levels, if a segregating site is heterozygous in $i$ and $j$ then $i$ and $j$ will share all $k$-mers between them.

To generalize beyond haploid organisms, we will replace the conversion factor of 2 in equation \ref{eq:haploid_diff2k} above with a general function $a(x)$:

\begin{equation} \label{eq:general_diff2k}
    |K_i \cup K_j| - |K_i \cap K_j| \leq a(x)\sum_{y<z}D_{yz}
\end{equation}

where $a(x)$ is a conversion factor that turns a number of pairwise differences into a number sample-exclusive $k$-mers as a function of $x$ and $\sum_{y<z}D_{yz}$ is all of the pairwise differences for any pair of haplotypes $y$ and $z$ in $i$ and $j$. The values that maximize $a(x)$, maintaining the validity of using a ``$\leq$'' sign in Equation \ref{eq:general_diff2k}, are as follows:

\begin{equation}
      a(x)=\begin{cases}
    \frac{2k}{x^2}, & \text{if $1 \leq x \leq 3$}.\\
    \frac{k}{2x-1}, & \text{if $x \geq 4$}.
  \end{cases}
\end{equation}

When the population is haploid ($x = 1$), Equation \ref{eq:general_diff2k} reduces to Equation \ref{eq:haploid_diff2k}. When $x =$ 2 or 3, $a(x)$ is maximized when $i$ and $j$ are homozygous for different alleles, creating $2k$ sample-exclusive $k$-mers for every 4 or 9 pairwise differences, respectively. However, when $x \geq 4$, the situation that maximizes $a(x)$ is one where the SNP exists on only one haplotype in $i$ or $j$, creating $k$ sample-exclusive $k$-mers for every $2x-1$ pairwise differences.

Now we will convert the right side of Equation \ref{eq:general_diff2k} into nucleotide diversity ($\pi$). Summing across all pairs of individuals gives:

\begin{equation}
    \sum_{i<j}|K_i \cup K_j| - |K_i \cap K_j| \leq a(x)\sum_{i<j}\sum_{y<z}D_{yz}
\end{equation}

Next, we convert the right-hand side into genome-wide average $\pi$ by dividing both sides by the number of pairwise haplotype comparisons \citep{korunes_pixy_2021}, which is the number of individuals sampled $n$ times ploidy $x$, choose 2:

\begin{equation}
    \frac{\sum_{i<j}|K_i \cup K_j| - |K_i \cap K_j|}{{nx\choose2}} \leq a(x) \left( \frac{\sum_{i<j}\sum_{y<z}D_{yz}}{{nx\choose2}} \right)
\end{equation}

\begin{equation}
    \frac{\sum_{i < j} |K_i \cup K_j| - |K_i \cap K_j|}{{nx\choose2}} \leq a(x)\pi
\end{equation}

Next, isolating $\pi$ on one side gives:

\begin{equation}\label{eq:ktopi}
    \frac{\sum_{i < j} |K_i \cup K_j| - |K_i \cap K_j|}{a(x){nx\choose2}} \leq \pi
\end{equation}

The intuition behind this formula is that, in a world where our assumptions are met, the number of $k$-mers that are not shared between a pair of samples is bounded by a multiple of $\pi$. The need to scale $\pi$ by $a(x)$ to get a bound reflects the fact that a given SNP can be captured by multiple $k$-mers, but the exact relationship between pairwise differences and sample-exclusive $k$-mers depends on ploidy. Note that we could covert the left-side of the equation \ref{eq:ktopi} to Jaccard dissmilarity (equation \ref{eq:jaccard}) if we divide both sides by $\sum_{i < j} |K_i \cup K_j|$.

Finally, we can substitute $\pi$ for the standard formula for the expected value of $\pi$ in a neutrally evolving population at equilibrium \citep{tajima_amount_1996}:

\begin{equation} \label{eq:expectedk}
    E \left[ \frac{\sum_{i < j} |K_i \cup K_j| - |K_i \cap K_j|}{a(x){nx\choose2}} \right] \leq \frac{2xN_e\mu}{1 + \frac{4}{3}2xN_e\mu}
\end{equation}

where $N_e$ is effective population size and $\mu$ is the mutation rate (probability of mutation per base pair per generation). The denominator on the right hand side of Equation \ref{eq:expectedk} ensures that the expected value of $\pi$ saturates as it approaches it's theoretical maximum of 0.75 \citep{tajima_amount_1996}. Altogether, equations \ref{eq:ktopi} and \ref{eq:expectedk} provide simple bounds for the average number of $k$-mers that differentiate a pair of samples in terms of $N_e$ and $\mu$.

\begin{figure}
    \centering
    \includegraphics[scale = 0.55]{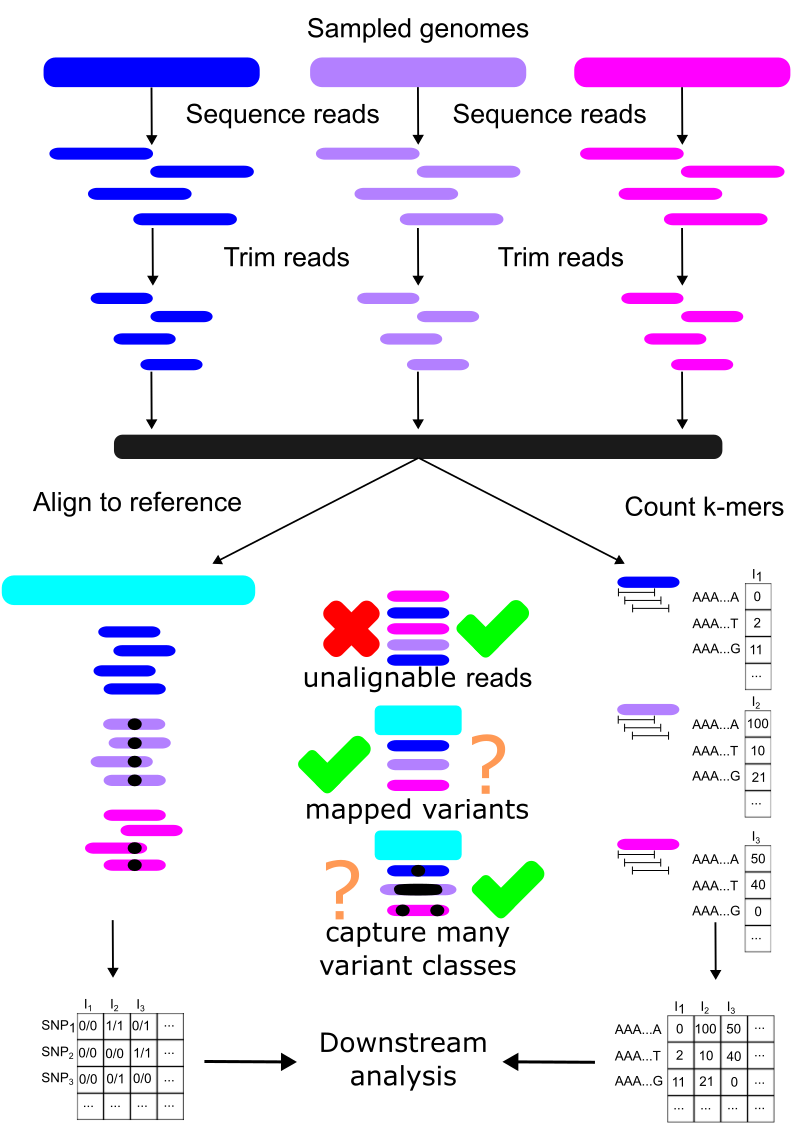}
    \caption{\textbf{Comparison between a typical SNP-calling workflow and a $k$-mer counting workflow.} A typical $k$-mer-based analysis begins the same as a SNP-based analysis: sequencing reads from sampled genomes followed by trimming and other quality control steps on the reads. The crucial difference comes down to whether the remaining reads are aligned to a reference sequence (discarding any unaligned reads) or are used for $k$-mer counting. Workflows may vary in terms of attempting to map $k$-mers to specific loci or calling variants other than SNPs (denoted with ``?''). The end result for each workflow is typically a matrix where each row is a different genomic variant, each column is a sample, and the elements represent the variant states. For SNPs, the standard variant call format notation is to use 0/0 to represent homozygous reference genotypes, 0/1 or 1/0 to represent heterozygous genotypes, and 1/1 to represent homozygous alternate genotypes. In $k$-mer based analyses the variant states are instead counts of how many $k$-mers of a particular type were observed in each sample.}
    \label{fig:kvs}
\end{figure}

\section*{Identifying variation with $k$-mers}

Many of the earlier studies using $k$-mers to identify variants focus on microbes, where (despite small genome sizes) high levels of diversity make alignment and choosing a reference genome difficult \citep{gardner_scalable_2010}. However, similar approaches have now been extended to other taxa. One class of approaches calls SNPs \textit{de novo} simply by comparing $k$-mers between samples and searching for pairs of $k$-mers that differ at their central basepair \citep{gardner_scalable_2010, gardner_when_2013, bedo_information_2016, li_kmer2snp_2022}. Another, more popular, class of methods relies on de Bruijn graphs, which are graphs where $k$-mers are nodes and two nodes are connected if they share $k - 1$ base pairs \citep{compeau_why_2011}. SNPs then appear as ``bubbles'' in these graphs and calling SNPs amounts to searching for these bubbles \citep{iqbal_novo_2012, leggett_identifying_2013, younsi_using_2015, uricaru_reference-free_2015, standage_kevlar_2019, gauthier_discosnp-rad_2020}. Finally, other approaches first identify $k$-mers that are present in all samples (i.e. ``anchor'' $k$-mers) and finds paths between anchor $k$-mers through either local alignment or traversing de Bruijn graphs \citep{audano_mapping-free_2018, kaplinski_katk_2021, aylward_pankmer_2023}. The main drawback to these approaches is that the relative positions of the resulting variant calls are usually unknown without further analysis (see Figure \ref{fig:kvs}).

Other methods can call variants with known relative positions by comparing $k$-mers to a database of previously-identified variants. For example, one can compare $k$-mers with a reference set of SNPs  \citep{shajii_fast_2016, pajuste_fastgt_2017, denti_malva_2019, shi_identifying_2023, chu_ntsm_2024} or insertions \citep{puurand_alumine_2019} to quickly genotype samples because SNPs often correspond to multiple unique $k$-mers. Or if a reference pangenome assembly is available, it's possible to use $k$-mers to infer the path through the pangenome corresponding to a given sample genome \citep{ebler_pangenome-based_2022, grytten_kage_2022, hantze_effects_2023}. However, similar to common alignment-based SNP calling practices, the called variants will be limited to variants present in the pangenome reference.

\section*{Measuring variation with $k$-mers}

After calling variants, quantifying levels of variation is a crucial step in many population genetics workflows. Standard approaches to this goal are to align sample sequences to a reference then calculate either (1) the average level of heterozygosity across sites or (2) the number of variants segregating in the sample. These are represented by Watterson's (Equation \ref{eq:watterson}; \citet{watterson_number_1975}) and Nei's (Equation \ref{eq:nei}; \citet{nei_mathematical_1979, nei_dna_1981}) estimators of diversity, respectively:

\begin{equation} \label{eq:watterson}
    \theta_w = \frac{S}{\sum_i^{n-1}\frac{1}{n}}
\end{equation}

\begin{equation} \label{eq:nei}
    \pi = (1 - \sum_i p_{i}^{2})(\frac{n}{n - 1})
\end{equation}
where $n$ is the number of sequences in the sample, $p_i$ is the frequency of the $i$th allele at a locus, and $S$ is the number of segregating sites at a locus. While $\theta_w$ is based on a discrete count of variants ($S$), $\pi$ is shaped by the allele frequencies of variants and will be higher if variants are common than if variants are rare (note that $\pi$ is more commonly rewritten in terms of the average number of differences between sequences \citep{korunes_pixy_2021}). 

Can analogous measures of variation be derived from $k$-mers? There are over 30 valid measures of genetic difference based on $k$-mer counts used in previous literature \citep{benoit_multiple_2016, zielezinski_alignment-free_2017, luczak_survey_2019, zielezinski_benchmarking_2019}. However, the three most common $k$-mer dissimilarity measures are arguably Jaccard dissimilarity (Equation \ref{eq:jaccard}; \citet{ondov_mash_2016}), Bray-Curtis dissimilarity (Equation \ref{eq:bray}; \citet{dubinkina_assessment_2016, benoit_simkamin_2020}), and cosine dissimilarity (Equation \ref{eq:cos}, \citet{choi_libra_2019}):

\begin{equation} \label{eq:jaccard}
    J(K_i,K_j) = 1 - \frac{K_i \cap K_j}{K_i \cup K_j}
\end{equation}

\begin{equation} \label{eq:bray}
    B(C_i,C_j) = 1 - 2\sum^{4^k}_{b = 1}\frac{\min(m_b(C_i), m_b(C_j)) }{m_b(C_i) + m_b(C_j)}
\end{equation}

\begin{equation} \label{eq:cos}
    C(C_i,C_j) = 1 - \frac{C_i \cdot C_j}{\|C_i\| \times \|C_j\|}
\end{equation}
where $k$ is the length of $k$-mers to be included in the comparison, $K_i$ and $K_j$ are the set of $k$-mers of length $k$ present in a set of reads $i$ and $j$, $C_i$ and $C_j$ are vectors of $k$-mer counts in a set of reads $i$ and $j$, and $m_i$ is a function that returns the relative frequency of the $b$th $k$-mer in a set (i.e. standardized such that $\sum^{4^k}_{b = 1} m_b(C_i)$ and $\sum^{4^k}_{b = 1} m_b(C_j)$ equal 1) . These measures work similarly to the classical $\pi$ and $\theta_w$ measures: the numerators are a measure of the number of sites that vary between individuals, while the denominators are a measure of sample size (here the number of $k$-mers rather than number of haplotypes). The main difference, however, is that $k$-mer-based measures of genetic dissimilarity are not directly interpretable in terms of mutations, like $\pi$ and $\theta_w$ can be with the assumption that each SNP represents one mutation \citep{Haubold_alignment-free_2011, Haubold_alignment-free_2012}. Any given $k$-mer may represent the combined presence of multiple mutations \citep{voichek_identifying_2020,blanca_statistics_2022} and any mutation can generate multiple new $k$-mers. Although this means that $k$-mer-based genetic dissimilarity measures are only proxies for the true mutational distance between individuals, they still effectively resolve relationships between lineages compared to alignment-based measures \citep{vanwallendael_alignment-free_2022}.

While Equations \ref{eq:jaccard}, \ref{eq:bray}, \ref{eq:cos} have all been successfully used to measure genetic dissimilarity between samples in past studies, the formulae highlight their benefits and drawbacks. First, Jaccard dissimilarity, perhaps the most commonly used $k$-mer dissimilarity metric \citep{ondov_mash_2016, ruperao_exploring_2023}, requires only knowing $k$-mer presence/absence patterns in samples instead of $k$-mer counts. It can take less memory to calculate than other $k$-mer-based dissimilarity measures. However, as a consequence it may not capture the effects of copy number variation and does not account for variation in coverage between samples, which affect whether a given $k$-mer is called as ``present'' in a sample \citep{vanwallendael_alignment-free_2022}. Approaches that measure $k$-mer counts, like Bray-Curtis dissimilarity (Equation \ref{eq:bray}) and cosine dissimilarity (Equation \ref{eq:cos}), can better account for these influences, but may require more memory for storing counts \citep{liu_unbiased_2017, choi_libra_2019}. To alleviate this memory problem, many approaches calculate approximate $k$-mer dissimilarity measures with a small subset of $k$-mers \citep{ondov_mash_2016, zhao_bindash_2019, benoit_simkamin_2020, Pellegrina_fast_2020}. An alternative approach would be to instead compress the $k$-mer counts into a smaller array, keeping information from more $k$-mers while simultaneously alleviating memory burdens \citep{melsted_efficient_2011}, but such approaches have not been used to calculate genetic dissimilarity before. The relationship between $k$-mer-based dissimilarity measures and $\pi$ is also rarely explored. While some studies have investigated the relationship between $\pi$ and Jaccard dissimilarity \citep{vanwallendael_alignment-free_2022}, other $k$-mer dissimilarity measures are possible and no studies to our knowledge have compared these approaches for populations of varying levels of diversity - a key determinant of whether alignment-based genotype calls are accurate \citep{cornish_comparison_2015, bush_genomic_2020}. In the following sections, we investigate the efficacy of compressed and uncompressed $k$-mer-based measures of variation at capturing the true pairwise diversity of simulated populations.

\subsection*{Testing the efficacy of $k$-mer measures of variation.}
We used simulations to investigate the relationship between the true value of $\pi$ and genetic diversity measured from $k$-mer based approaches from simulated sequencing reads. 

\subsubsection*{Simulations}
We simulated a neutrally evolving 100kb segment of the \textit{Arabidopsis thaliana} genome. We first forward simulated 300 neutrally evolving populations with SLiM 3 \citep{haller_slim_2019}. Each population consisted of 100 individuals simulated for 1000 generations with a uniform recombination rate of $10^{-8}$. To vary the diversity across simulations we varied the mutation rate between $10^{-6}$ and $2*10^{-4}$. For each simulation, we tracked the ancestry through tree-sequence recording which records the geneological history of all samples \citep{slimTreeRecording}. From these trees, we generated sequences based on the \textit{Arabidopsis thaliana} genome (chromosome 1 at positions 4,185,001-4,285,000) \citep{Kent2002-xv}. We took a sub-sample of 10 individuals from the tips of the trees and randomly assigned nucleotides to each SNP in the sample using msprime version 1.2.0 and tskit version 0.5.6 \citep{baumdicker2022efficient}. From the sub-tree of the sampled genomes, we recorded the exact number of true average pair-wise differences ($\pi_t$) across the sample of 10 individuals (20 chromosomes) using msprime \citep{baumdicker2022efficient}. 

\subsubsection*{Generating $k$-mers}
To test the performance of the $k$-mer measures on unaligned reads, we simulated reads for each genome using an Illumina read simulator, InSilicoSeq 2.0.0 \citep{gourle_simulating_2019}. We varied the read count to later investigate the effect of coverage as described below. We generated two sets of reads for each individual at coverages of 10x and 30x. After simulating reads, we counted $k$-mers within the reads using KMC3 \citep{deorowicz_kmc_2015, kokot_kmc_2017}. We generated $k$-mer count vectors with $k=10, 20, 30,$ and $40$ for each individual. We used a threshold-based approach to adjust the $k$-mer vectors to reduce the effects of sequencing errors; any $k$-mer count below the threshold value was set to zero for a particular sample before any dissimilarity calculations. For a threshold of 5, only $k$-mers with counts of 5 or more were considered when calculating the difference between two or more groups of $k$-mer counts. We present data with a threshold of 5, but note that a threshold of 0 is qualitatively similar with dissimilarity scores being slightly higher overall. This practice of filtering out low-coverage $k$-mers is analogous to the common practice of filtering out SNPs below a given minor allele frequency threshold \citep{asif_gwas_2021}. After $k$-mer counting, we calculated the genetic dissimilarity of populations with the Bray-Curtis (Equation \ref{eq:bray}) and cosine dissimilarity (Equation \ref{eq:cos})) measures.

\subsubsection*{The effect of $k$ on $k$-mer similarity metrics}

\begin{figure}
  \centering
  \includegraphics[width=\linewidth]{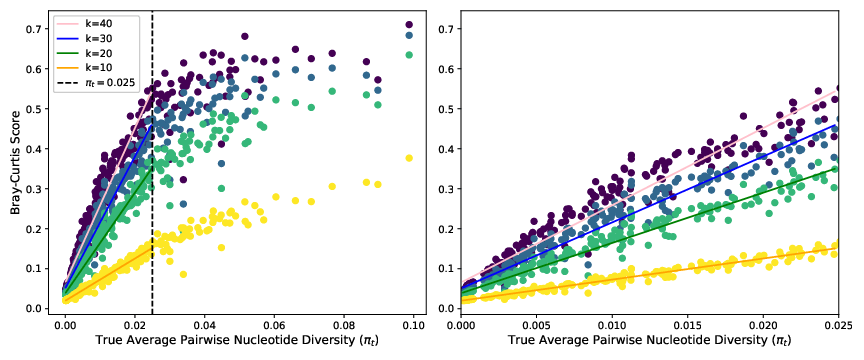}
  \caption{Bray-Curtis dissimilarity calculated from simulated reads with a coverage of 30 where each point represents a sample of 20 chromosomes. Displayed are 10-mers (yellow), 20-mers (green), 30-mers (blue), and 40-mers (purple) each with a linear regression line for $\pi_t$ between 0 and 0.025. The scores for populations with a $\pi_t$ less than 0.025 is shown to the right.}
  \label{fig:BCCov30Kmers}
\end{figure}

Figure \ref{fig:BCCov30Kmers} shows the effect of $k$ on the Bray-Curtis score with 30x coverage. We observe a plateau in the scores when diversity exceeds $\pi\approx2.5\%$. This plateau occurs because, when diversity is high, single nucleotide polymorphisms cause most of the $k$-mers to be different between two samples. This effect is especially true with high $k$ values as one variant sampled in only one read can appear in up to $2k$ $k$-mers if it is sampled away from the edges of a read. The elevated number of $k$-mers at high diversity means that it is harder to interpret differences in dissimilarity measures between samples above this threshold, but this problem can be mitigated by using a lower $k$ value, such as $k$=10 when the expected diversity in a sample is high. If $\pi_t$ is expected to be below 0.025, larger $k$ values can have higher precision and capture more unique $k$-mers in individual samples, better estimating true diversity.

While the data presented in Figure \ref{fig:BCCov30Kmers} were simulated with a coverage of 30x, Figure \ref{fig:bc_coverage} shows that a coverage of 10x results in qualitatively similar scores when $k$=30. While coverage affects the precision of the measures, the overall trends and relative rankings between simulations remain consistent. 

\begin{figure}
  \centering
  \includegraphics[width=.8\linewidth]{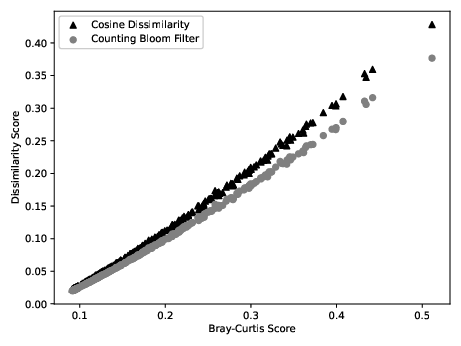}
  \caption{Cosine dissimilarity (triangle) and counting bloom filter scores (circle) from 10-mers of simulated reads with a coverage 30 compared directly against the Bray-Curtis score of each simulated set of 10 individuals (points).}
  \label{fig:AllThree}
\end{figure}

\subsubsection*{The Counting Bloom filter approach to comparing genomes}
\begin{center}
\begin{figure}[!ht]
  \includegraphics[width=\linewidth]{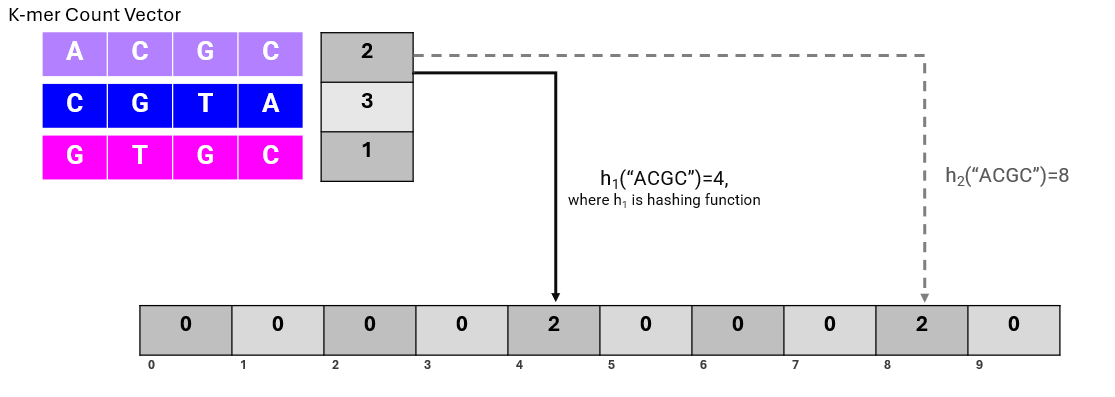}
  \caption{A schematic of a Counting Bloom Filter where the first 4-mer, ‘ACGC’ is hashed. The hash function $h_1$ outputs 4 and $h_2$ outputs 8. Both positions in the CBF array are incremented by 2 as that is the count of ‘ACGC’ in the given $k$-mer count vector}
  \label{fig:CBFDiagram}
\end{figure}
\end{center}

Bloom filters are a data structure that can compress a $k$-mer count vector into a smaller array \citep{melsted_efficient_2011}. We can compute the dissimilarity of two compressed vectors using less memory and with increased efficiency as fewer entries are compared. Counting bloom filters (CBFs) are modifications of bloom filters \citep{fan2000summarycache}, and are used in alternative $k$-mer count methods and error correction in sequencing data \citep{melsted_efficient_2011, roy2014turtle, shi2010quality}. CBFs compress the $k$-mer vectors through several hash functions that map to a smaller array. Counting bloom filters will increase the $k$-mer count in the position as opposed to setting the count to one if a $k$-mer has been mapped there (see Figure \ref{fig:CBFDiagram}). Because of the hash-functions and collisions, false positives are possible; here a ``false positive'' means that two $k$-mers of different sequences may map to the same locations in the vector (i.e. have a hash collision). However, identical mapping of two different $k$-mers is not a concern for relative comparison of diversity, since the same hash functions are used for each sample so collisions are consistent across samples. Collisions will cause diversity to be underestimated, but we can reduce collisions by increasing the number of hash-functions used in the process \citep{melsted_efficient_2011}.

A CBF gives a standardized way to compare across species/experiments if the vectors for $k$-mer counts are set to the same size and the same hash functions are used to generate the vectors across data sets. We can also perform the same cosine dissimilarity measure on the CBF vector that we use directly on the counts of $k$-mers. The cosine dissimilarity measure on the raw $k$-mer count vectors has the same $R^2$ to the compressed CBF vectors ($0.97$) when comparing the scores to $\pi_t$. Figure \ref{fig:AllThree} shows that the cosine dissimilarity using the counting bloom filter is qualitatively similar to the cosine dissimilarity with the raw $k$-mer counts. This similarity and the high $R^2$ mean that the predictability of $\pi_t$ is maintained while using the smaller, more manageable CBF data structure. Therefore we can reliably use the CBF data structure as opposed to the raw $k$-mer counts to calculate the cosine dissimilarity of two samples.

The memory usage of a $k$-mers vector for a sample with 10x where $k=30$ under our simulations is around 4.5 MB per sample. The memory usage is decreased to 0.02 MB when compressed to a 10,000 element array of unsigned 16-bit integers. Therefore, storing and using $k$-mer counts for measuring diversity scale much better under the CBF data structure.

\subsubsection*{The effect of array size on CBF measures of diversity}

\begin{figure}[hbt!]
\centering
  \includegraphics[width=\linewidth]{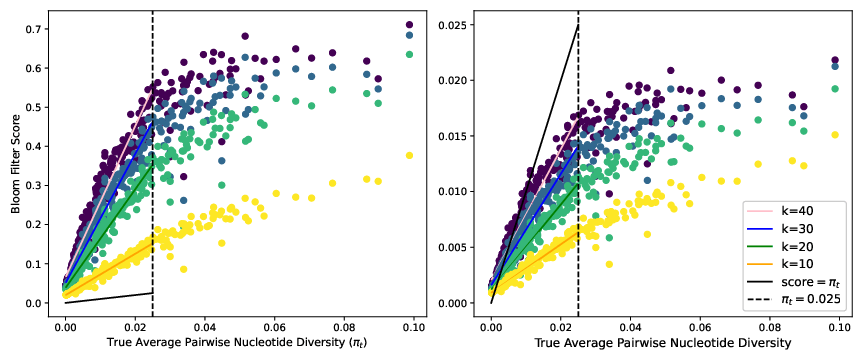}
  \caption{A comparison of cosine similarity scores of CBFs with an array size of 20 million (left) and 10,000 (right). The black line shows the 1-to-1 mapping of $\pi_t$ to the dissimilarity score. Note the difference in the scales of the scores on the y-axes, and that the right panel shows points that are closer to the 1-to-1 line.}
  \label{fig:cbf_array}
\end{figure}

We found that the size of the CBF array scales the cosine dissimilarity score and keeps the rank of scores very similar as the array size changes. Since rank is maintained, we can take advantage of much smaller arrays without losing much information about the relative diversity scores between samples. Smaller arrays allow for less memory usage, and faster computing of cosine dissimilarity. Supplementary Figure \ref{fig:cbf_array} compares the scores of a CBF array of size 20 million and 10 thousand on the same simulations. We observe that the smaller array adjusts the scores down while maintaining the same rank of scores. For example, these results show that if a population has the lowest cosine score with the 20 million length array, it remains among the lowest scores with a 10 thousand length array. This result means that it may be possible to adjust the vector size 
 such that the cosine dissimilarity scores are similar in magnitude and scale to the true pairwise nucleotide diversity which can be useful for prediction of the score without alignment, and can make interpretation simpler. However, the viability of this approach still needs to be investigated across different species and genomic contexts, it is likely there is not a `one size fits all' array size that would scale cosine dissimilarity to match $\pi$ universally. Our simulations only consider SNPs, neutral evolution, and the \textit{Arabidopsis thaliana} genome. It is still left to determine the effects of other mutations types such as insertions, deletions, and inversions as well as other evolution types. Therefore we should avoid the potential pitfalls of interpreting a scaled value directly as $\pi$ unless future work shows that to be appropriate.

\section*{Explaining variation with $k$-mers}

A common first step to understanding the evolutionary forces shaping variation in populations involves quantifying differentiation between populations. Patterns of differentiation in SNP genotypes are often summarized and plotted using dimension reduction techniques, such as principal component analysis (PCA) \citep{novembre_interpreting_2008}. $k$-mer genotypes are also amenable to PCA and recover the same differentiation patterns as SNP-based PCA \citep{liu_unbiased_2017, murray_kwip_2017, rahman_association_2018, ho_intraspecific_2019, hrytsenko_determining_2022}. Applying dimensional reduction to $k$-mers can even differentiate species \citep{rosen_metagenome_2008, aflitos_cnidaria_2015, bernard_recapitulating_2016, linard_rapid_2019, bodde_high-resolution_2022}, something that can be difficult to do with SNPs without MSA. Interestingly, $k$-mers from repetitive sequences may not always differentiate populations that are clearly differentiated in terms of their SNP genotypes \citep{renny-byfield_repetitive_2020}, suggesting that identifying the set of $k$-mers that best differentiate populations could be an important avenue for future research. Furthermore, while we are aware of one study in yeast that estimates admixture proportions from $k$-mers \citep{shi_alignment-_2024}, more investigation is needed to see if this approach works for other species.

Besides polymorphism between populations, $k$-mers are also useful for explaining polymorphism patterns across genomes in terms of specific evolutionary forces. For instance, the $k$-mer profile of a given locus is often predictive of the local recombination rate \citep{liu_sequence-dependent_2012, Haubold_alignment-free_2013, Haubold_alignment-free_2014, frenkel_organizational_2016, al_maruf_irspot-sf_2019} and the local mutation rate \citep{aggarwala_expanded_2016, carlson_extremely_2018, bethune_method_2022, adams_regularized_2023, beichman_evolution_2023, liu_structural_2023}. $k$-mers can capture information about recombination and mutation because these processes often associate with specific functional DNA motifs \citep{myers_common_2008, ruzicka_dna_2017}. However, $k$-mers are also intrinsically sensitive to sequence changes and while there are some existing methods to identify new mutations \citep{nordstrom_mutation_2013, ho_intraspecific_2019} and sites of recombination \citep{Fletcher_AFLAP_2021} solely from $k$-mers, further development is needed. Being able to predict fine scale variation in mutation and recombination rates solely from the $k$-mer profile of a reference sequence would be extremely beneficial because these processes frequently confound scans for sites of selection \citep{huber_detecting_2016}. However, this would require further study of the relationship between $k$-mers and mutation/recombination rates across wider ranges of species.

$k$-mers also offer opportunities to explore patterns of selection. One general assumption in $k$-mer literature is if a $k$-mer is shared across all individuals of a population or species then it is potentially under selection to be conserved \citep{bernard_recapitulating_2016, aylward_pankmer_2023}. In contrast, a $k$-mer that is not present in any individual is possibly selected against (although this assumes that $k$ is sufficiently small such that the total possible $k$-mer space is small, as might be true when analysing small amino acid-derived $k$-mers)  \citep{Georgakopoulos-Soares_absent_2021}. However, it is unlikely that selection is solely responsible for patterns of $k$-mer sharing across individuals or species, so more rigorous approaches are needed. One potentially more rigorous approach would be to investigate $k$-mers that differentiate populations as candidates for loci underlying local adaptation. Although, we are not aware of any papers that employ this specific approach, identifying group-specific $k$-mers is a very common and useful practice, with one example being the assembly of sex-specific sequences \citep{Akagi_y-chromosomeencoded_2014, Ou_ngs-based_2017, Liao_genomic_2020, Neves_male_2020, mehrab_efficient_2021, wu_inferring_2021, Behrens_sex_2022, Fong_evolutionary_2023, Lichilin_no_2023}. An alternative approach would be to march through a deBruijn graph \citep{aylward_pankmer_2023} to find strings of $k$-mers with selective sweep-like patterns, mainly low diversity, high linkage disequilibrium between $k$-mers, an excess of rare $k$-mers \citep{alachiotis_raisd_2018}, high differentiation between populations \citep{zhong_hard_2022}, and high haplotype homozygosity \citep{klassmann_detecting_2022} which could be determined based on $k$-mer copy number \citep{vurture_genomescope_2017,ranallo-benavidez_genomescope_2020}.

There are only a few model-oriented studies that compare the $k$-mer spectrum observed in a genome to a neutral expectation, mostly in the context of detecting selection on transcription factor binding motifs \citep{gerland_selection_2002, ke_positive_2008, raijman_evolution_2008, yeang_quantifying_2010, gyorgy_competition_2023}. The basic idea behind these approaches is to first derive a neutral substitution model - which describes the probability of one nucleotide being substituted for another in a neutrally evolving sequence \citep{ke_positive_2008, raijman_evolution_2008} - then measure deviations from that neutral model according to the presence/absence of particular $k$-mers in a genome. Although these models are able to detect selection on $k$-mers that exist in multiple places across a genome, as is the case with binding motifs, it is unclear whether they could be applied to $k$-mers that represent unique genomic sequences. 

\section*{Challenges in investigating $k$-mers} 

Despite the wide potential of $k$-mers to be useful for population genetic studies, there are three important limitations to keep in mind for most $k$-mer investigations. First, interpreting the biology of specific $k$-mers is often a challenge. One option is to take the candidate $k$-mers identified from an analysis and either align the $k$-mers themselves, the reads containing said $k$-mers, or assembled reads containing said $k$-mers to a reference genome \citep{voichek_identifying_2020} or a database of known sequences and motifs. While this is effective at pinpointing concrete type of variants at play in a system, it partially defeats the purpose of using $k$-mers in the first place because $k$-mers of interest may not be present in a reference genome. Second, since each sequencing error can generate up to $k$ erroneous $k$-mers and general practice is to discard reads with many unique $k$-mers to reduce the effect of error rates \citep{zimin_masurca_2013}, analysis of $k$-mers typically requires datasets with high coverage ($>$10x) and low sequencing error rates. These criteria can potentially exclude long-read datasets \citep{vurture_genomescope_2017} or reduced-representation datasets where genomes are sequenced at lower coverage to cut costs. Approaches that apply $k$-mers in lower coverage or higher error-rate datasets are needed. Third, $k$-mer-based analyses can frequently involve hundreds of millions or billions of unique $k$-mers. Storing and processing this many $k$-mers at once is often not feasible even on high performance computing systems. Two approaches to solving this issue are to subset one's $k$-mers \citep{ondov_mash_2016, benoit_simkamin_2020, zhao_bindash_2019, yi_kssd_2021}, or, as we discuss here, compress vectors of $k$-mer counts into an array of smaller size \citep{melsted_efficient_2011}. However, these solutions are usually developed for specific contexts, such as measuring genomic dissimilarity, and may not be applicable to every population genetic analysis. Future studies need to carefully consider common approaches to decreasing the disk burden of $k$-mer-based analyses and should explore new potential solutions.

\section*{Conclusions}

Current literature demonstrates that $k$-mers are useful for identifying, measuring, and explaining variation within populations without performing MSA. Through our own simulations of neutrally evolving populations, we find that $k$-mer dissimilarity reliably scales with nucleotide diversity and $k$-mer matrices can be compressed with minimal loss of dissimilarity information. However, further development is needed to make $k$-mers amenable to a wider array of population genetic tasks, especially the identification of selected loci and population structure. Further developing alignment-free approaches to population genetics tasks will ultimately address bottlenecks in the analysis of pangenomes and provide valuable complements to common alignment-based approaches.

\section*{Data availability}

Scripts for simulation and data analysis can be found at: \url{https://github.com/williarj/kmers2024}.

\section*{Acknowledgments}

We would like to thank the members of the Josephs lab for feedback on earlier drafts of this work.

\section*{Funding}

This work was funded by a National Institutes of Health grant (R35 GM142829) to E.B.J., an Integrated Training Model in Plant And Computational Sciences Fellowship (National Science Foundation: DGE-1828149) to M.D.R., a Plant Biotechnology for Health and Sustainability Fellowship (National Institute Of General Medical Sciences of the National Institutes of Health : T32-GM110523) to M.D.R., and a Michigan State University Institute for Cyber-Enabled Research Cloud Computing Fellowship to M.D.R. The content of this article is solely the responsibility of the authors and does not necessarily represent the official views of the National Institutes of Health.

\section*{Conflicts of interest}

None declared

\section*{Supplementary material}

\subsection*{The effect of coverage on $k$-mer measures}

\begin{figure}[!ht]
\centering
  \includegraphics[width=0.9\linewidth]{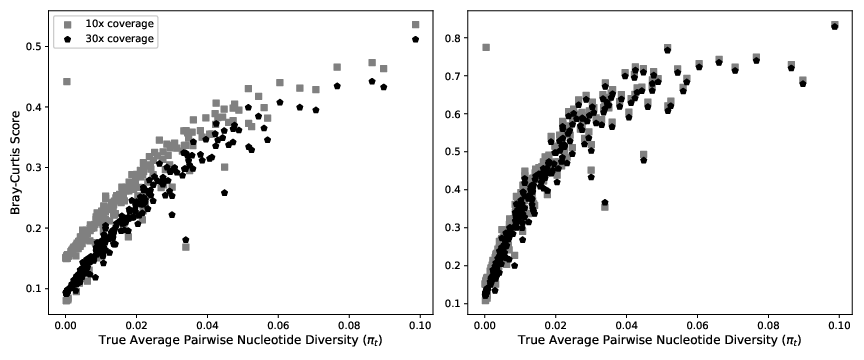}
  \caption{Bray-Curtis scores calculated with 10-mers (left) and 30-mers (right). For each sample, reads with 10x coverage (gray/square) and 30x coverage (black/pentagon) were simulated. After simulating reads, the $k$-mers were counted and then the Bray-Curtis score is calculated. Each point represents one sample with a specific $k$ and coverage.}
  \label{fig:bc_coverage}
\end{figure}

We observe is Figure \ref{fig:bc_coverage} a complex intereaction between coverage and the resulting Bray-Curtis score depending on the length of $k$. With 30-mers, there is very little difference in the Bray-Curtis score. However, 10-mers are more affected by changes in coverage. The overall trend and ranks of the scores remain, but the Bray-Curtis score is generally higher with 10x coverage.

\newpage

\subsection*{Adjusting array size of Counting Bloom Filter}

\bibliography{library,olivia_refs}

\bibliographystyle{abbrvnat}
\setcitestyle{authoryear, open={(},close={)}}

\end{document}